\documentclass[12pt]{article}
\usepackage{epsf}
\usepackage[dvips]{color}
\usepackage{exscale}
\usepackage[dvips]{graphicx}
\usepackage{amsmath}
\usepackage{amssymb}
\usepackage{latexsym}

\begin{document}
\setlength{\baselineskip}{18pt}
\begin{titlepage}

\vspace*{1.2cm}
\begin{center}
{\Large\bf Magnetic dipole moment and keV neutrino dark matter}
\end{center}
\lineskip .75em
\vskip 1.5cm

\begin{center}
{\large 
Chao-Qiang Geng$^{a,b,}$\footnote[1]{E-mail:
\tt geng@phys.nthu.edu.tw} 
and 
Ryo Takahashi$^{a,}$\footnote[2]{E-mail: 
\tt ryo.takahasi88@gmail.com}
}\\

\vspace{1cm}

$^a${\it Department of Physics, National Tsing Hua University, Hsinchu 300, 
Taiwan}

$^b${\it Physics Division, National Center for Theoretical Sciences, Hsinchu 
300, Taiwan}

\vspace*{10mm}
{\bf Abstract}\\[5mm]
{\parbox{13cm}{\hspace{5mm}
We study magnetic dipole moments of right-handed neutrinos in a keV neutrino dark matter 
model. This model is a simple extension of the standard model with only  right-handed neutrinos and 
a pair of charged particles added. One of the right-handed neutrinos is the candidate of dark matter
with a keV mass. Some bounds  on the dark matter magnetic dipole moment
and model parameters are obtained from  cosmological observations.
}}
\end{center}
\end{titlepage}

\section{Introduction}

The evidences from neutrino experiments have established the neutrino 
oscillation phenomenon. The experimental data can be explained by flavor 
mixings of three (active) neutrinos, and oscillation probabilities are 
described by three generation mixing angles and two mass squared differences. 
The presence of non-vanishing neutrino masses means a necessity 
of physics beyond the standard model (SM). Furthermore, the smallness of neutrino 
mass squared differences compared with  the charged fermion masses in the SM is one of 
striking properties of neutrinos. The type-I seesaw mechanism~\cite{seesaw} is 
the most promising approach to explain such smallness of neutrino masses. In this
 mechanism, the sterile (right-handed) neutrinos are added to the SM. 
Since these sterile neutrinos are Majorana particles, they can have Majorana masses, which 
violate the lepton number. Consequently, the heavy enough Majorana masses of the sterile 
neutrinos can lead to tiny active neutrino masses.

On the other hand, the elucidation of the origin of dark matter 
(DM)~\cite{Zwicky:1933gu}, which governs about $23\%$ of the energy density of 
the Universe~\cite{Komatsu:2010fb}, is also one of the most important goals in
particle physics today. Recently, a large number of DM models have been discussed in the 
literature (e.g. see~\cite{Garrett:2010hd} for a recent review, and references 
therein). Among them, one of interesting candidates for DM is a sterile 
neutrino. In particular, models with three sterile neutrinos whose masses are 
below the EW scale have been proposed 
in~\cite{Gouvea:2005er,Asaka:2005an,Asaka:2005pn}. Note that some astrophysical
 data possibly support the existence of sterile 
neutrinos~\cite{Kusenko:2009up}. In addition to a candidate for DM, the sterile
 neutrino can also play a role in other cosmological issues, such as the origin 
of the baryon asymmetry of the Universe (BAU). For instance, relatively heavy 
sterile neutrinos in the type-I seesaw mechanism can generate the BAU via
 leptogenesis~\cite{RIFP-641}. The split seesaw mechanism~\cite{Kusenko:2010ik}
 can give a hierarchical mass spectrum of sterile neutrinos, which can 
incorporate the usual leptogenesis with a keV sterile neutrino DM 
scenario.\footnote{See also~\cite{Adulpravitchai:2011rq} for discussions of 
the realistic flavor mixing in the mechanism.} 
A possible mass spectrum of sterile neutrinos to explain
 the BAU~\cite{hep-ph/0203194}, 
 DM, and MiniBooNE/LSND 
oscillation anomalies~\cite{Aguilar:2001ty}, as well as realize the tiny active 
neutrino masses, has been proposed in~\cite{Chen:2011ai}. 
Clearly, the nature of 
neutrinos would be a key to find physics beyond the SM and understand some 
cosmological issues.

In this letter, we focus on the magnetic property of neutrinos. 
In particular, we concentrate on a simple 
extension of the SM~\cite{Aparici:2009mj}, in which the right-handed neutrino 
magnetic moment can be generated by the interaction between new charged particles and sterile neutrinos. 
Note that the 
induced magnetic interaction could result in some interesting consequences for 
cosmology and high energy physics.
%

\section{Magnetic dipole moments}
\subsection{Active neutrinos}

By introducing three generations of the right-handed neutrinos in the SM, the 
Yukawa interactions are given by
 \begin{eqnarray}
  \mathcal{L}=\mathcal{L}_{{\rm SM}}-(y_\nu\bar{L}\nu_R\Phi+h.c.),
  \label{Lag}
 \end{eqnarray}
where $L$, $\nu_R$, and $\Phi$ are the left-handed lepton doublet, right-handed
 neutrino, and the SM Higgs, respectively. 
 The Dirac 
neutrino mass is given by $M_D=y_\nu v$, where $v$ is  the vacuum expectation value (VEV) of the 
Higgs.

The Dirac neutrino can have a magnetic dipole moment  induced by 
radiative corrections~\cite{Gunn:1978gr} as
 \begin{eqnarray}
  \mu_{\nu_i}=\frac{3eG_F}{8\sqrt{2}\pi^2}m_{\nu_i}\simeq3\times10^{-19}
              \left(\frac{m_\nu}{1\mbox{ eV}}\right)\mu_B, \label{mag}
 \end{eqnarray}
where $G_F$ is the Fermi constant, $m_{\nu_i}$ are the corresponding (Dirac) 
neutrino mass eigenvalues, and $\mu_B$ is the Bohr magneton, given by
 \begin{eqnarray}
  \mu_B=\frac{e}{2m_e}=5.79\times10^{-9}\mbox{ eV}\cdot\mbox{Gauss}^{-1}
      =1.93\times10^{-11}\mbox{ e cm}.
 \end{eqnarray}
In Eq.~\eqref{mag}, we have assumed $m_\nu\simeq m_{\nu_i}$ as
the typical  neutrino mass scale. The current upper bounds on the 
neutrino magnetic moments for three flavors are given by the Borexino 
experiment as~\cite{Studenikin:2008bd}
 \begin{eqnarray}
  \mu_{\nu_e}<5.4\times10^{-11}\mu_B,~~~
  \mu_{\nu_\mu}<1.5\times10^{-10}\mu_B,~~~
  \mu_{\nu_\tau}<1.9\times10^{-10}\mu_B,
 \end{eqnarray}
respectively. Note that a stronger bound on the typical neutrino dipole moment ($\mu_\nu$) is inferred 
in~\cite{Raffelt:1990pj} from an estimate of effects on the core mass of the red 
giants at the helium flash as
 \begin{eqnarray}
  \mu_\nu<3\times10^{-12}\mu_B~~~\mbox{ with }~~~
  \mu_\nu^2\equiv\sum_{i,j=1}^3(|\mu_{ij}|^2+|\epsilon_{ij}|^2), \label{rg}
 \end{eqnarray}
where $\mu_{ij}$ and $\epsilon_{ij}$ are the elements of the magnetic and electric dipole 
matrices, respectively.

Here, we mention that the transition magnetic moment, which is relevant to 
$\nu_i\rightarrow\nu_j+\gamma$, may exist for both Dirac and Majorana neutrino 
cases. Explicitly, one has 
\cite{petcov}
 \begin{eqnarray}
  \mu_{ij}^D=\frac{3eG_F}{32\sqrt{2}\pi^2}(m_{\nu_i}+m_{\nu_j})
                    \sum_{\alpha=e,\mu,\tau}U_{j\alpha}^\dagger U_{\alpha i}
                    \left(\frac{m_\alpha}{m_W}\right)^2,
 \end{eqnarray}
for the Dirac neutrinos, where $U$ and $m_\alpha$ are the 
Pontecorvo-Maki-Nakagawa-Sakata (PMNS) matrix and the corresponding charged 
lepton masses, respectively. If the neutrinos are Majorana particles, one can 
only have a flavor changing dipole operator,
 \begin{eqnarray}
  \mathcal{L}_{\text{int}}
   =\mu_{ij}^M\nu_iC^{-1}\sigma_{\mu\nu}\nu_jF^{\mu\nu}+h.c.,
 \end{eqnarray}
where $\nu_i$ are active neutrinos,\footnote{They describe the left-handed 
neutrinos as $\nu_{L\alpha}=U_{\alpha i}\nu_i+\theta_{\alpha I}N_I^c$ with the 
left-right mixing angles $\theta_{\alpha I}\equiv(y_\nu)_{\alpha I}v/M_I$ and 
other mass eigenstates $N_I$ (so-called sterile neutrinos)  almost corresponding
 to the right-handed neutrinos, $i.e.$ $N_I\simeq\nu_{RI}$. Throughout this letter, 
indices $i,j=1\sim3$ and $I,J=1\sim3$ stand for generations of active and 
sterile neutrinos, respectively.} leading to the transition magnetic moments~\cite{pal}:
 \begin{eqnarray}
  \mu_{ij}^M=\frac{3eG_F}{16\sqrt{2}\pi^2}(m_{\nu_i}+m_{\nu_j})
             \sum_{\alpha=e,\mu,\tau}\mbox{Im}\left[U_{j\alpha}^\dagger 
             U_{\alpha i}\left(\frac{m_\alpha}{m_W}\right)^2\right],
 \end{eqnarray}
for the Majorana neutrinos. 

\subsection{Right-handed neutrinos}

We now consider the magnetic moments of the right-handed neutrinos. If the 
neutrinos are Majorana particles, one can have the Majorana mass terms for the 
right-handed neutrinos, which violate the lepton number,
with the Lagrangian, given by
 \begin{eqnarray}
  \mathcal{L}_{{\rm Majorana}}=-\frac{M_R}{2}\overline{\nu_{R}^c}\nu_R. 
  \label{Majorana}
 \end{eqnarray}
It is well known that the Lagrangians in Eqs.~\eqref{Lag} and \eqref{Majorana} give 
light active neutrino masses through the seesaw mechanism~\cite{seesaw},
  \begin{eqnarray}
  M_\nu=-M_D^TM_R^{-1}M_D, \label{seesaw}
 \end{eqnarray} 
after integrating out the heavy right-handed neutrinos. 
Note that 
$y_\nu$, $M_R$,
$M_D$ and  $M_\nu$ are all $3\times3$ matrices. 

Since the right-handed (sterile)  neutrinos are Majorana particles, only the transition magnetic moments can be induced,
described by the following flavor changing dipole operator,
 \begin{eqnarray}
  \mathcal{L}_{{\rm int}}=\mu_{IJ}^NN_IC^{-1}\sigma_{\mu\nu}N_JF^{\mu\nu}+h.c..
 \end{eqnarray}
There are some models to induce the magnetic moments of right-handed (sterile) 
neutrinos.

\subsection{
Neutrino dark matter model}

Recently, Aparici, Santamaria, and Wudka (ASW) have proposed a 
model~\cite{Aparici:2009mj} which enlarges the SM by adding a negatively 
charged scalar, $\omega$, and one negatively charged vector-like fermion, $E$, 
with non-vanishing hypercharges, $Y(\omega)=-1$ and $Y(E)=-1$, in addition to 
the right-handed neutrinos. When one imposes a discrete symmetry, which affects
 only $\omega$ and $E$ as $\omega\rightarrow -\omega$ and $E\rightarrow-E$, the
 relevant Lagrangian allowed by the SM gauge and additional discrete symmetries
 is 
 \begin{eqnarray}
  \mathcal{L}_{{\rm ASW}} 
   &=& \mathcal{L}_{{\rm SM}}+\mathcal{L}_{{\rm Majorana}}+\mathcal{L}_K
       -\mathcal{L}_Y-V, \label{ASW} \\
  \mathcal{L}_K           
   &=& D_\mu\omega^\dagger D^\mu\omega+i\bar{E}D\hspace{-2.5mm}/\hspace{0.5mm}E
       +i\bar{\nu}_R\partial\hspace{-2.4mm}/\hspace{0.5mm}\nu_R-M_E\bar{E}E, 
       \label{LK} \\
  \mathcal{L}_Y
   &=& y_\nu\bar{L}\nu_R\Phi+h\bar{\nu}_RE\omega^++h.c., 
\end{eqnarray}
where  $V$ 
is the scalar potential for the SM Higgs and $\omega$.

In this model, 1-loop diagrams involving $\omega$ and $E$ contribute to the magnetic 
moments of the right-handed neutrinos. The same interactions 
also  give rise to  contributions 
to the right-handed neutrino Majorana masses through the operator 
$\xi(\Phi^\dagger\Phi)\overline{\nu_R^c}\nu_R$. An invisible Higgs decay 
through the interaction has been  discussed in~\cite{Aparici:2009mj}. 
Moreover, it has been  pointed out that the new charged particles 
can be produced at the CERN LHC experiment through the Drell-Yan process 
because of their charged properties if they are light enough. We further 
investigate DM properties of this kind of the model.

We now proceed with DM in the model.
It is known that one of interesting neutrino DM models is the keV sterile 
neutrino DM model (e.g. see~\cite{Asaka:2005an,Asaka:2005pn,deGouvea:2005er}).\footnote{See also \cite{vega} and \cite{itoh}
 for general discussions on DM properties with the keV mass and neutrino energy loss in stellar interiors, respectively.
} 
In this scenario, the lightest sterile neutrino with the keV  mass is a 
decaying DM candidate. To be DM, the lifetime of the lightest sterile neutrino 
should be larger than the age of the Universe. The lightest sterile neutrino 
can radiatively decay into a photon ($\gamma$) and an active neutrino ($\nu_i$)
through the 
left-right mixing. 

Since we have new interactions which generate the right-handed neutrino 
magnetic moments, the lightest sterile 
neutrino can radiatively decay into $\gamma$ and $\nu_i$ with the decay width,
 given by
 \begin{eqnarray}
  \Gamma_{N_1\rightarrow\nu_i\gamma}^{\rm mag}
  =\frac{(M_1^2-m_{\nu_i}^2)^3}{8\pi M_1^3}|\mu_{1i}|^2
  \simeq\frac{M_1^3}{8\pi}|\mu_{1i}|^2, \label{gammaIi}
 \end{eqnarray}
where $|\mu_{1i}|$ denotes the magnetic moment
and $M_1$ is the mass of the (keV) sterile 
neutrino ($N_1$).
Here, the active neutrino mass has been neglected in the second equality of Eq.~(\ref{gammaIi}). 
On the other hand, the 
keV sterile neutrino DM model also has a constraint from its decay into $\gamma$ and $\nu_i$
through the gauge boson and charged lepton loops with 
the left-right mixing angle. The decay width is given by
 \begin{eqnarray}
  \Gamma_{N_1\rightarrow\nu_i\gamma}
  =\frac{9\alpha G_F^2}{1024\pi^4}\sin^2(2\theta_1)M_1^5
  \simeq5.5\times10^{-22}\theta_1^2\left(\frac{M_1}{\mbox{keV}}\right)^5
        \mbox{ s}^{-1},
 \end{eqnarray}
where $\theta_1\equiv\sum_{\alpha=e,\mu,\tau}(y_\nu)_{\alpha1}v/M_{1}$. 
Clearly, both decay mechanisms could produce
a narrow line in the X-ray back 
ground~\cite{Boyarsky:2009ix,NewAdded,0709.2301}.
As a result, for the latter case,
  the left-right mixing angle 
is restricted as $\theta_1^2\lesssim1.8\times10^{-5}(\mbox{keV}/M_1)^5$,  
equivalently 
$\Gamma_{N_1\rightarrow\nu_i\gamma}\lesssim9.9\times10^{-27}\mbox{ s}^{-1}$. 
For the former, it is reasonable to
 impose a bound 
$\Gamma_{N_1\rightarrow\nu_i\gamma}\lesssim(10^{-28}-10^{-26})\mbox{ s}^{-1}$ 
in a region of the emission photon energy $0.5\mbox{ keV}\leq E_\gamma\leq 12$ 
keV given in \cite{0709.2301} on 
$\Gamma_{N_1\rightarrow\nu_i\gamma}^{\rm mag}$. The emission photon energy is 
related with the decaying sterile neutrino mass as $E_\gamma=M_1/2$. For 
$\Gamma_{N_1\rightarrow\nu_i\gamma}^{\rm mag}\lesssim10^{-28}\mbox{ s}^{-1}$, 
one obtains
 \begin{eqnarray}
  |\mu_{1i}|\lesssim3.89\times10^{-16}\mu_B, \label{new}
 \end{eqnarray} 
where $M_1=5$ keV has been 
used.\footnote{$\Gamma_{N_1\rightarrow\nu_i\gamma}\lesssim10^{-26}\mbox{ 
s}^{-1}$ is also allowed for $M_1\simeq24$ keV. In this case, a more severe 
bound $|\mu_{1i}|\lesssim3.70\times10^{-16}\mu_B$ can be derived.} It is seen that
 the constraint in Eq.~(\ref{new}) on the neutrino magnetic moment is much stronger 
than the one from the consideration of the red giants  in Eq.~(\ref{rg}). Note 
that Eq.~(\ref{rg}) is obtained from the discussion of the plasmon decay into 
neutrinos where the masses of neutrinos are lower than 
$\mathcal{O}(\mbox{keV})$. Therefore, once the sterile neutrinos  have 
magnetic interactions mediated by new particles, the keV sterile neutrino DM 
scenario should satisfy the  constraint in Eq.~(\ref{new}), which is 
model-independent~\cite{Aparici:2009mj}, rather than the one from the red giants.

We now investigate the neutrino magnetic moment in a model-dependent way. The
 magnetic moment $|\mu_{1i}|$ induced from the model in Eq.~(\ref{ASW}) is 
calculated as
 \begin{eqnarray}
  |\mu_{1i}|&=&      \frac{g'f(r)}{2(4\pi)^2M_E}\sum_{J=2,3}
                     \sum_{\alpha=e,\mu,\tau}
                     \mbox{Im}[h_1^\ast h_J\theta_{J\alpha}U_{\alpha i}], 
                     \label{mu-dep} \\
  f(r)      &\equiv& \frac{1}{1-r}+\frac{r}{(1-r)^2}\log(r), \label{fr}
 \end{eqnarray}
for the case of $M_1\ll M_E$ and $M_\omega$ with $r\equiv M_\omega^2/M_E^2$. 
Here, the active neutrino as the final state is converted from the internal 
sterile state $N_J\simeq\nu_{RJ}$ $(J=2,3)$ through the corresponding 
left-right mixing $\theta_{J\alpha}$. Since the Majorana neutrinos can only 
have the transition magnetic moments, the sum of $J$ is performed for $J=2$ and
 $3$. The external momenta and masses can be neglected  as 
in~\cite{Aparici:2009mj}. 


Two of three sterile neutrinos can generically play a role to realize the 
active neutrino mass scales through the seesaw mechanism in the keV sterile 
neutrino DM model, e.g.~\cite{Asaka:2005an}. Therefore, the left-right mixing 
angle for the corresponding generations can be described by the typical active 
neutrino mass scale $m_\nu$ and two heavier sterile neutrino mass scales 
$M_{2,3}$, given by $\theta_{J\alpha}=\sqrt{m_\nu/M_{2,3}}$. On the 
other hand, since the Yukawa coupling of the lightest sterile neutrino to the 
left-handed lepton doublet and SM Higgs should be tiny, the sterile neutrino DM
 with the keV mass is not responsible for the active neutrino mass scales. Because 
of this smallness of the Yukawa coupling, the keV sterile neutrino cannot be 
in the equilibrium even at a high temperature. This feature is crucial for 
the various production mechanisms of the keV sterile neutrino DM with the correct 
abundance~\cite{Kusenko:2010ik,hep-ph/9303287,astro-ph/9810076,0609081}.


We now explicitly examine
a specific and economical model~\cite{Asaka:2005an,Asaka:2005pn} 
with right-handed neutrinos and new charged particles  as an 
example. In this model, one of heavier sterile neutrinos is in the 
thermal equilibrium before the sphaleron process becomes 
inactive~\cite{akhmedov}. When the Yukawa coupling of 
the remaining heavier sterile neutrino is naively estimated as 
$(y_{\nu2})^2\sim\sqrt{\Delta m_{\rm sol}}M_2/v^2\sim\mathcal{O}(10^{-15})$, 
the sterile neutrino is out of equilibrium at the time without the sphaleron process.
The $2\leftrightarrow2$ interactions among the right-handed neutrinos and new charged 
particles, such as the scalar exchange $\nu_RE\leftrightarrow\nu_RE$ interaction, 
are important for the condition of the non-equilibrium of DM. The rates of those new interactions are 
described by the new Yukawa couplings $h_I$ given in Eq.~(\ref{LK}), where $I$ 
denotes the generation of the right-handed neutrinos. 
Note that these Yukawa couplings do not affect the active neutrino masses. When 
$|h_I|^2\lesssim\mathcal{O}(10^{-14})$, the corresponding sterile neutrino is 
out-of-equiribrium at the time when the spharelon process becomes ineffective. 
Under these discussions, we impose $\Gamma_{N_1\rightarrow\nu_i\gamma}^{\rm 
mag}\lesssim10^{-28}\mbox{ s}^{-1}$ on Eq.~(\ref{new}) with Eqs.~(\ref{mu-dep}) 
and (\ref{fr}). Then, we obtain a constraint on the model parameter as
 \begin{eqnarray}
  M_E\geq 24.3\mbox{ MeV}\,,
  \label{EqConstraint}
 \end{eqnarray} 
where we have taken that $g'=0.35$, $f(r)=1/2$, 
$\theta_{J\alpha}=\sqrt{m_\nu/M_{2,3}}$, $m_\nu=0.01$ eV, $M_{2,3}=10$ GeV, and
 Im$[h_1h_3U_{\alpha i}]=5\times10^{-9}$.  Note that these values can satisfy the above 
conditions in the keV sterile neutrino DM model realizing the BAU via the 
oscillation of the heavier sterile neutrino with a mass spectrum of
$(M_1,M_2,M_3)=(\mbox{keV},\mathcal{O}(1-10)\mbox{ GeV},\mathcal{O}(1-10)\mbox{
 GeV})$. Note also that $f(r)\rightarrow1/2$ if $M_\omega/M_E\rightarrow1$.
 It is clear the  constraint in Eq.~(\ref{EqConstraint}) is much weaker than that from high energy experiments 
in the presence of the new charged particles.
In other models of the BAU, the constraint on the model parameters becomes
 weaker because of the largeness of the heavier sterile neutrino masses.

\section{Summary}

We have investigated the magnetic dipole moments in the keV sterile neutrino DM
 model. In this DM model, the lightest sterile neutrino with the keV scale mass
 is a decaying DM candidate with its lifetime greater than the age of the 
Universe. Since the width of the radiative DM decay into a photon and an active
 neutrino is constrained by X-ray observations, we have obtained a 
model-independent constraint on the magnetic interactions, leading to 
$|\mu_{1i}|\lesssim3.89\times10^{-16}\mu_B$ for $M_1=5$ keV, which is stronger 
than the bound from the consideration of the plasmon decay in the red giants. 
We have also studied the magnetic dipole moment in a model-dependent way. 
Explicitly, 
the same condition from the X-ray observations gives a constraint of $M_E\gtrsim 24.3\mbox{ MeV}$ in the model of baryogenesis 
from the heavier right-handed neutrino oscillation.

\subsection*{Acknowledgement}

This work was supported in part by the National Science Council of 
Taiwan under Grant No. NSC-98-2112-M-007-008-MY3 and National Center for 
Theoretical Sciences, Taiwan.



\begin{thebibliography}{99}
\bibitem{seesaw}
P.~Minkowski,
Phys.~Lett. {\bf B67} (1977) 421;
T.~Yanagida, in Proceedings of the Workshop on Unified Theories
and Baryon Number in the Universe, eds.\ O.~Sawada and A.~Sugamoto
(KEK report 79-18, 1979); M.~Gell-Mann, P.~Ramond and R.~Slansky, in
Supergravity, eds.\ P.~van~Nieuwenhuizen and D.Z.~Freedman
(North Holland, Amsterdam, 1979); R.~N.~Mohapatra and G.~Senjanovic, 
 Phys.\ Rev.\ Lett.\  {\bf 44} (1980) 912;
J.~Schechter and J.~W.~F.~Valle,
Phys.\ Rev.\ D {\bf 22} (1980) 2227; Phys.\ Rev.\ D {\bf 25} (1982) 774.

\bibitem{Zwicky:1933gu}
  F.~Zwicky,
  Helv.\ Phys.\ Acta {\bf 6} (1933) 110.

\bibitem{Komatsu:2010fb}
  E.~Komatsu {\it et al.}  [WMAP Collaboration],
  Astrophys.\ J.\ Suppl.\  {\bf 192} (2011) 18.

\bibitem{Garrett:2010hd}
  G.~Bertone, D.~Hooper and J.~Silk,
  Phys.\ Rept.\  {\bf 405} (2005) 279;
  G.~D'Amico, M.~Kamionkowski and K.~Sigurdson,
  arXiv:0907.1912 [astro-ph.CO];
  K.~Garrett and G.~Duda,
  Adv.\ Astron.\  {\bf 2011} (2011) 968283.

\bibitem{Gouvea:2005er}
  A.~de Gouvea,
  Phys.\ Rev.\ D {\bf 72} (2005) 033005.  

\bibitem{Asaka:2005an}
  T.~Asaka, S.~Blanchet and M.~Shaposhnikov,
  Phys.\ Lett.\  B {\bf 631} (2005) 151.

\bibitem{Asaka:2005pn}
  T.~Asaka and M.~Shaposhnikov,
  Phys.\ Lett.\  B {\bf 620} (2005) 17.

\bibitem{Kusenko:2009up}
  A.~Kusenko and G.~Segre,
  Phys.\ Lett.\  B {\bf 396} (1997) 197;
  G.~M.~Fuller, A.~Kusenko, I.~Mocioiu and S.~Pascoli,
  Phys.\ Rev.\  D {\bf 68} (2003) 103002;
  P.~L.~Biermann and A.~Kusenko,
  Phys.\ Rev.\ Lett.\  {\bf 96} (2006) 091301; 
  M.~Mapelli, A.~Ferrara and E.~Pierpaoli,
  Mon.\ Not.\ Roy.\ Astron.\ Soc.\  {\bf 369} (2006) 1719; 
  J.~Stasielak, P.~L.~Biermann and A.~Kusenko,
  Astrophys.\ J.\  {\bf 654} (2007) 290;
  E.~Ripamonti, M.~Mapelli and A.~Ferrara,
  Mon.\ Not.\ Roy.\ Astron.\ Soc.\  {\bf 375} (2007) 1399;
  F.~Munyaneza and P.~L.~Biermann,
  Astron.\ Astrophys.\  {\bf 458} (2006) L9;
  J.~Stasielak, P.~L.~Biermann and A.~Kusenko,
  Acta Phys.\ Polon.\  B {\bf 38} (2007) 3869;
  A.~Kusenko, B.~P.~Mandal and A.~Mukherjee,
  Phys.\ Rev.\  D {\bf 77} (2008) 123009;
  M.~Loewenstein and A.~Kusenko,
  Astrophys.\ J.\  {\bf 714} (2010) 652;
  D.~A.~Prokhorov and J.~Silk,
  arXiv:1001.0215 [astro-ph.HE];
  M.~Lovell, V.~Eke, C.~Frenk, L.~Gao, A.~Jenkins, T.~Theuns, J.~Wang and A.~Boyarsky {\it et al.},
  arXiv:1104.2929 [astro-ph.CO].

\bibitem{RIFP-641}
  M.~Fukugita and T.~Yanagida,
  Phys.\ Lett.\ B\ {\bf 174} (1986) 45.  

\bibitem{Kusenko:2010ik}
  A.~Kusenko, F.~Takahashi and T.~T.~Yanagida,
  Phys.\ Lett.\  B {\bf 693} (2010) 144.

\bibitem{Adulpravitchai:2011rq}
  A.~Adulpravitchai and R.~Takahashi,
  JHEP {\bf 1109} (2011) 127.

\bibitem{hep-ph/0203194}
  M.~Fukugita and T.~Yanagida,
  Phys.\ Rev.\ Lett.\ \ {\bf 89} (2002) 131602.

\bibitem{Aguilar:2001ty}
  A.~Aguilar {\it et al.}  [LSND Collaboration],
  Phys.\ Rev.\  D {\bf 64} (2001) 112007;
  A.~A.~Aguilar-Arevalo {\it et al.}  [The MiniBooNE Collaboration],
  Phys.\ Rev.\ Lett.\  {\bf 105} (2010) 181801.
  
\bibitem{Chen:2011ai}
  C.~S.~Chen and R.~Takahashi,
  arXiv:1112.2102 [hep-ph].

\bibitem{Aparici:2009mj}
  A.~Aparici, A.~Santamaria and J.~Wudka,
  J.\ Phys.\ G {\bf 37} (2010) 075012.
  
\bibitem{Gunn:1978gr}
  W.~J.~Marciano and A.~I.~Sanda,
  Phys.\ Lett.\ B {\bf 67} (1977) 303;
  B.~W.~Lee and R.~E.~Shrock,
  Phys.\ Rev.\  D {\bf 16} (1977) 1444;
  J.~E.~Gunn, B.~W.~Lee, I.~Lerche, D.~N.~Schramm and G.~Steigman,
  Astrophys.\ J.\  {\bf 223} (1978) 1015.
  K.~Fujikawa and R.~Shrock,
  Phys.\ Rev.\ Lett.\  {\bf 45} (1980) 963.

\bibitem{Studenikin:2008bd}
  D.~Montanino, M.~Picariello and J.~Pulido,
  Phys.\ Rev.\  D {\bf 77} (2008) 093011;
  C.~Arpesella {\it et al.}  [The Borexino Collaboration],
  Borexino Data,''
  Phys.\ Rev.\ Lett.\  {\bf 101} (2008) 091302;
  A.~Studenikin,
  Nucl.\ Phys.\ Proc.\ Suppl.\  {\bf 188} (2009) 220.

\bibitem{Raffelt:1990pj}
  G.~G.~Raffelt,
  Phys.\ Rev.\ Lett.\  {\bf 64} (1990) 2856.

\bibitem{petcov}
S.~T.~Petcov,
Sov.\ J.\ Nucl.\ Phys.\  {\bf 25} (1977) 340   [Yad.\ Fiz.\  {\bf 25} (1977) 641]   [Erratum-ibid.\  {\bf 25} (1977) 698]   [Erratum-ibid.\  {\bf 25} (1977) 1336].

\bibitem{pal}
P.~B.~Pal and L.~Wolfenstein,
Phys.\ Rev.\ D {\bf 25} (1982) 766.

\bibitem{deGouvea:2005er}
  T.~Asaka, M.~Shaposhnikov and A.~Kusenko,
  Phys.\ Lett.\ B {\bf 638} (2006) 401;
  M.~Shaposhnikov,
  Nucl.\ Phys.\  B {\bf 763} (2007) 49;
  T.~Asaka, M.~Laine and M.~Shaposhnikov,
  JHEP {\bf 0606} (2006) 053;
  T.~Asaka, M.~Laine and M.~Shaposhnikov,
  JHEP {\bf 0701} (2007) 091;
  A.~Boyarsky, J.~Lesgourgues, O.~Ruchayskiy and M.~Viel,
  Phys.\ Rev.\ Lett.\  {\bf 102} (2009) 201304;  
   X.~G.~He, T.~Li and W.~Liao,
  Phys.\ Rev.\ D {\bf 81}, 033006 (2010);
  F.~Bezrukov, H.~Hettmansperger and M.~Lindner,
  Phys.\ Rev.\  D {\bf 81} (2010) 085032;
   W.~Liao,
  Phys.\ Rev.\ D {\bf 82}, 073001 (2010)
  M.~Lindner, A.~Merle and V.~Niro,
  JCAP {\bf 1101} (2011) 034;
  A.~Merle and V.~Niro,
  JCAP {\bf 1107} (2011) 023;
    J.~Barry, W.~Rodejohann and H.~Zhang,
  arXiv:1110.6382 [hep-ph];
  T.~Araki and Y.~F.~Li,
  arXiv:1112.5819 [hep-ph];
  A.~Merle,
  arXiv:1201.0881 [hep-ph].

\bibitem{vega}
H.~J.~de Vega and N.~G.~Sanchez,
Mon.\ Not.\ Roy.\ Astron.\ Soc.\  {\bf 404} (2010) 885;  
H.~J.~de Vega, P.~Salucci and N.~G.~Sanchez,
arXiv:1004.1908 [astro-ph.CO];
H.~J.~de Vega and N.~G.~Sanchez, 
Int.\ J.\ Mod.\ Phys.\ A {\bf 26} (2011) 1057.  

\bibitem{itoh}
N.~Itoh, H.~Hayashi, A.~Nishikawa, and Y.~Kohyama, 
ApJS 102 (1996) 411.


\bibitem{Boyarsky:2009ix}
  A.~D.~Dolgov and S.~H.~Hansen,
  Astropart.\ Phys.\  {\bf 16} (2002) 339;  
  K.~Abazajian, G.~M.~Fuller and W.~H.~Tucker,
  Astrophys.\ J.\  {\bf 562} (2001) 593;  
  A.~Boyarsky, A.~Neronov, O.~Ruchayskiy and M.~Shaposhnikov,
  Mon.\ Not.\ Roy.\ Astron.\ Soc.\  {\bf 370} (2006) 213;  
  A.~Boyarsky, A.~Neronov, O.~Ruchayskiy and M.~Shaposhnikov,
  JETP Lett.\  {\bf 83} (2006) 133;
  A.~Boyarsky, A.~Neronov, O.~Ruchayskiy and M.~Shaposhnikov,
  Phys.\ Rev.\ D {\bf 74} (2006) 103506;  
  A.~Boyarsky, A.~Neronov, O.~Ruchayskiy, M.~Shaposhnikov and I.~Tkachev,
  Phys.\ Rev.\ Lett.\  {\bf 97} (2006) 261302;  
  S.~Riemer-Sorensen, S.~H.~Hansen and K.~Pedersen,
  Astrophys.\ J.\  {\bf 644} (2006) L33;  
  C.~R.~Watson, J.~F.~Beacom, H.~Yuksel and T.~P.~Walker,
  Phys.\ Rev.\ D {\bf 74} (2006) 033009;  
  S.~Riemer-Sorensen, K.~Pedersen, S.~H.~Hansen and H.~Dahle,
  Phys.\ Rev.\ D {\bf 76} (2007) 043524;  
  A.~Boyarsky, J.~Nevalainen and O.~Ruchayskiy,
  Astron.\ Astrophys.\  {\bf 471} (2007) 51;  
  K.~N.~Abazajian, M.~Markevitch, S.~M.~Koushiappas and R.~C.~Hickox,
  Phys.\ Rev.\ D {\bf 75} (2007) 063511;
  A.~Boyarsky, O.~Ruchayskiy and M.~Markevitch,
  Astrophys.\ J.\  {\bf 673} (2008) 752;  
  A.~Boyarsky, J.~W.~den Herder, A.~Neronov and O.~Ruchayskiy,
  Astropart.\ Phys.\  {\bf 28} (2007) 303;  
    A.~Boyarsky, O.~Ruchayskiy, M.~Shaposhnikov,
  Ann.\ Rev.\ Nucl.\ Part.\ Sci.\  {\bf 59 } (2009)  191-214.


\bibitem{NewAdded}
  M.~Loewenstein, A.~Kusenko and P.~L.~Biermann,
Astrophys.\ J.\  {\bf 700} (2009) 426.


\bibitem{0709.2301}
  A.~Boyarsky, D.~Iakubovskyi, O.~Ruchayskiy and V.~Savchenko,
  Mon.\ Not.\ Roy.\ Astron.\ Soc.\  {\bf 387} (2008) 1361.

\bibitem{hep-ph/9303287} 
  S.~Dodelson and L.~M.~Widrow,
  Phys.\ Rev.\ Lett.\  {\bf 72} (1994) 17.  

\bibitem{astro-ph/9810076}  
   X.~-D.~Shi and G.~M.~Fuller,
   Phys.\ Rev.\ Lett.\  {\bf 82} (1999) 2832.  

\bibitem{0609081}
  A.~Kusenko,
  Phys.\ Rev.\ Lett.\  {\bf 97} (2006) 241301;  
  K.~Petraki and A.~Kusenko,
  Phys.\ Rev.\ D {\bf 77} (2008) 065014;  
  A.~Kusenko,
  Phys.\ Rept.\  {\bf 481} (2009) 1;  
  I.~M.~Shoemaker, K.~Petraki and A.~Kusenko,
  JHEP {\bf 1009} (2010) 060.  

\bibitem{akhmedov}
E.~K.~Akhmedov, V.~A.~Rubakov and A.~Y.~.Smirnov,
  Phys.\ Rev.\ Lett.\  {\bf 81} (1998) 1359.  

\bibitem{hep-ph/0309342}
  A.~Pilaftsis and T.~E.~J.~Underwood,
  Nucl.\ Phys.\ B\ {\bf 692} (2004) 303.  
\end{thebibliography}
\end{document}